\documentclass{aa}
\usepackage{graphicx}
\usepackage{times}
\begin{document}
\title{Line profile variability in the spectra of Oef stars:\\ II. HD\,192281, HD\,14442 and HD\,14434\thanks{Based on observations collected at the Observatoire de Haute Provence, France.}}
\author{M.\ De Becker \and G.\ Rauw\thanks{Research Associate FNRS (Belgium)}} 
\offprints{M. De Becker}
\institute{Institut d'Astrophysique et de G\'eophysique, Universit\'e de Li\`ege,
17, All\'ee du 6 Ao\^ut, B5c, B-4000 Sart Tilman, Belgium}

\date{Received date / Accepted date}
\authorrunning{M.\,De Becker \& G.\,Rauw }
\titlerunning{Line profile variability in the spectra of Oef stars:\\ part II.}

\abstract{We present the very first analysis of the spectroscopic variability of the three rapidly rotating Oef stars HD\,192281 (O5(ef)), HD\,14442 (O5.5ef) and HD\,14434 (O6.5(ef)). Radial velocities of the \ion{He}{ii} $\lambda$ 4541 line reveal no evidence of binarity on time scales of a few days, or from one year to the next, for any of the targets. The \ion{He}{ii} $\lambda$ 4686 double-peaked emission and, to some extent, the H$\beta$ absorption line display significant profile variability in the spectra of all three stars. Data gathered during different observing runs spread over six years reveal a rather stable time scale for HD\,192281 and HD\,14442,  whereas the variability pattern changes significantly from one year to the other. The case of HD\,14434 is less clear as no obvious time scale emerges from our analysis. In a temptative way to interpret this variability, stellar rotation remains a possible clock for HD\,192281 and HD\,14442. However, currently available models addressing stellar rotation fail to explain some crucial aspects of the observed variability behaviour, which appear to be even more complex in the case of HD\,14434.
}
\maketitle

\section{Introduction}

Monitoring of the spectroscopic variability has been used extensively to study the structure of the stellar winds of hot luminous stars (e.g. Fullerton et al.\,\cite{Ful}; Massa et al.\,\cite{Massa}; Kaper et al.\,\cite{Kap1}, \cite{Kap2}). Some objects were found to display cyclical variations in their stellar wind lines on time scales of a few days, possibly related to the rotational period of the underlying star. Large scale co-rotating structures in the stellar winds due to photospheric features (magnetic fields, non-radial pulsations,...) could possibly account for this variability (e.g. Kaper et al.\,\cite{Kap2}; Rauw et al.\,\cite{Rauw}). In order to quantify the importance of different processes that could contribute to this variability, more stars should be investigated. The objects of the scarce category of so-called Oef stars (Conti \& Leep\,\cite{CL}) appear a priori as promising targets for such an investigation since the double peaked morphology of the \ion{He}{ii} $\lambda$ 4686 emission line has been attributed to the effect of fast rotation.\\
The best studied Oef stars are $\zeta$\,Pup (O4ef) and $\lambda$\,Cep (O6ef). The two stars display strong spectral variability partly attributed to the effect of rotation (Moffat \& Michaud \cite{MM}; Howarth et al.\,\cite{How}; Kaper et al.\,\cite{Kap2}). In the case of $\zeta$\,Pup, Eversberg et al. (\cite{ELM}) further reported stochastic modulations of the \ion{He}{ii} $\lambda$ 4686 line explained by clumps distributed across the wind. These authors also confirmed the existence of spectral modulations on a time scale compatible with the rotation period, near the core of the \ion{He}{ii} $\lambda$ 4686 line.\\
Recently, Rauw et al.\,(\cite{RDV}, Paper I) analysed the complex behaviour of the \ion{He}{ii} $\lambda$ 4686 line in the spectrum of BD\,+60$^\circ$\,2522 (O6.5ef). They suggested that rotation was not the only clock ruling this variability, and that possibly non-radial pulsations were responsible for some short term variability. In this paper, we report the results of an intensive study of three other members of the small group of Oef stars, i.e HD\,192281, HD\,14442 and HD\,14434.

\begin{table}
\caption{Summary of some properties of HD\,192281, HD\,14442 and HD\,14434. References are the following: (1) Conti \& Leep (\cite{CL}), (2) Walborn (\cite{Wal1}), (3) Walborn (\cite{Wal2}), (4) this study, (5) Bieging et al.\,(\cite{BAC}), (6) Conti \& Burnichon (\cite{CB}), (7) Leitherer (\cite{Lei}), (8) Kendall et al.\,(\cite{Ken}), (9) Gies \& Bolton (\cite{GB}), (10) Humphreys (\cite{Hum}), (11) Conti \& Ebbets (\cite{CE}).\label{gen}}
\begin{center}
\begin{tabular}{l c c c}
\hline
\hline
	& HD\,192281 & HD\,14442 & HD\,14434 \\
\hline
Sp.T. & O5.5(ef)$^1$ & O6ef$^1$ & O6.5(ef)$^{1,4}$\\
	& O5Vn((f))$^{2}$ & O5n(f)p$^{3}$ & O5.5Vn((f))p$^2$\\
	& O5(ef)$^{4}$ & O5.5ef$^{4}$ & \\
$\log$(T$_\mathrm{eff}$) & 4.67$^{5}$ & 4.62$^{6}$ & 4.60$^{6}$\\
M$_\ast$ (M$_\odot$) & 70$^{7}$ & 44-54$^{8,a}$ & 44-52$^{8,a}$\\
 & 93$^{5}$ &  & \\
R (R{$_\odot$}) & 15$^{7}$ & 7.8-16.6$^{8,b}$ & 10.0-12.6$^{8,b}$\\
 & 19$^{5}$ &  & \\
d (kpc) & 1.78$^{9}$ & 2.88$^{8}$ & 3.00$^{8}$\\
Association & Cyg\,OB8$^{10}$ & Per\,OB1$^{10,c}$ & Per\,OB1$^{10,c}$\\
V$_{rot}\sin$i (km\,s$^{-1}$) & 270$^{11}$ & 273$^{11}$ & 400$^{11}$\\
\hline
\end{tabular}
\begin{list}{}{}
\item[$^{\mathrm{a}}$] The two values quoted are respectively the spectroscopic and the evolutionary masses.
\item[$^{\mathrm{b}}$] The first value of the radius is calculated for an assumed value of the distance modulus, and the second one is derived from the spectral type.
\item[$^{\mathrm{c}}$] The membership to this association is uncertain (see Kendall et al.\,\cite{Ken2})
\end{list}
\end{center}
\end{table}

HD\,192281 (V\,=\,7.6) is an O5(ef) star of the Cyg OB8 association. On the basis of its large radial velocity quoted in the literature, this star was considered as a runaway candidate by Gies \& Bolton (\cite{GB}), but the results of their study did not confirm this status. The radial velocity of HD\,192281 was also studied by Barannikov (\cite{Bar}) who argued that it was a binary with a 5.48 day orbital period, but this result has, so far, never been confirmed.\\
HD\,14442 (V\,=\,9.2) and HD\,14434 (V\,=\,8.5) are respectively O5.5ef and O6.5(ef) stars which were believed to belong to the Per OB1 association. On the one hand, HD\,14442 has been classified as nitrogen enriched (ON) by Bisiacchi et al. (\cite{Bis}). On the other hand, HD\,14434 may have a normal nitrogen spectrum (Schild \& Berthet \cite{SB}). Both stars have been reported as blue straggler candidates by Kendall et al. (\cite{Ken}) who considered they could have followed a quasi-homogeneous evolutionary track. However, in a subsequent study, Kendall et al. (\cite{Ken2}) found no CNO abundance anomalies in the spectrum of HD\,14442 and HD\,14434. These authors therefore rejected the blue straggler status of these stars and questioned their membership in Per OB1, suggesting that their distance could be larger than that of this association.\\
The three stars have a large projected rotational velocity: Conti \& Ebbets (\cite{CE}) reported V$_{rot}\sin$i values of 270, 273 and 400 km\,s$^{-1}$ for HD\,192281, HD\,14442 and HD\,14434 respectively. Very few studies have been performed on these three stars. In Table\,\ref{gen}, we summarize the information regarding their fundamental parameters that we have gathered from the literature. Different spectral types have been assigned to these stars and some of them are given in Table\,\ref{gen} for information. We shall briefly return to this point in Sect.\,\ref{spec}.\\

This paper is the second (and last) one of a series devoted to the line profile variability of Oef stars. It describes the first line profile variability study performed on HD\,192281, HD\,14442 and HD\,14434. In Sect.\,\ref{obs}., we present the observations and the data reduction. Section\,\ref{spec}. discusses the mean spectrum and the spectral type of our three target stars. Radial velocities and equivalent widths are discussed in Sect.\,\ref{rvew}. Section\,\ref{lpv} describes the results of our line profile variability analysis, and Sect.\,\ref{disc} consists in a discussion of our results and in an attempt of a consistent interpretation. The conclusions and prospects for future work are finally given in Sect.\,\ref{conc}.

\section {Observations and data reduction \label{obs}}
Spectroscopic observations of our three targets were collected during seven observing campaigns between September 1998 and October 2003 at the Observatoire de Haute-Provence (OHP). All observations were carried out with the Aur\'elie spectrograph fed by the 1.52 m telescope (Gillet et al.\, \cite{Gil}). For the 1998 and 1999 missions, Aur\'elie was equipped with a Thomson TH7832 linear array with a pixel size of 13 $\mu$m. From 2000 on, the detector was replaced by a 2048$\times$1024 CCD EEV 42-20\#3, with a pixel size of 13.5 $\mu$m squared.\\
An overview of the different campaigns is displayed in Table\,\ref{camp}. Data reduction followed the same approach as described in Paper I. Table\,\ref{camp} also specifies for each campaign which star was observed, as well as the spectral range covered during the run. In the following, we will merge the data collected during the two 1999 campaigns (July and August) into one single set which will be called Summer 1999, because of the small number of spectra taken during these two individual runs.  

\begin{table*}
\caption{Observing campaigns for the study of HD\,192281, HD\,14442 and HD\,14434. The last three columns specify the number of spectra obtained for each star during a specific observing run. \label{camp}}
\begin{center}
\begin{tabular}{l c c c c c c c}
\hline
\hline
Campaign & Detector & Spectral range & Recipr. disp. & Res. power & HD\,192281 & HD\,14442 & HD\,14434\\
 & & (\AA) & (\AA\,mm$^{-1}$) & & & & \\
\hline
September 1998 & TH7832 & 4455 - 4890 & 16 & 8000 & 6 & 0 & 0 \\
July 1999 & TH7832 & 4100 - 4950 & 33 & 4000 & 7 & 5 & 0 \\
August 1999 & TH7832 & 4100 - 4950 & 33 & 4000 & 4 & 7 & 0 \\
September 2000 & CCD & 4455 - 4905 & 16 & 8000 & 0 & 9 & 0 \\
September 2001 & CCD & 4455 - 4905 & 16 & 8000 & 16 & 12 & 0 \\
September 2002 & CCD & 4455 - 4905 & 16 & 8000 & 10 & 9 & 11 \\
October 2003 & CCD & 4455 - 4905 & 16 & 8000 & 7 & 6 & 19 \\
\hline
\end{tabular}
\end{center}
\end{table*}

\section{Mean spectrum \label{spec}}

\begin{figure*}
\begin{center}
\resizebox{17.5cm}{12.0cm}{\includegraphics{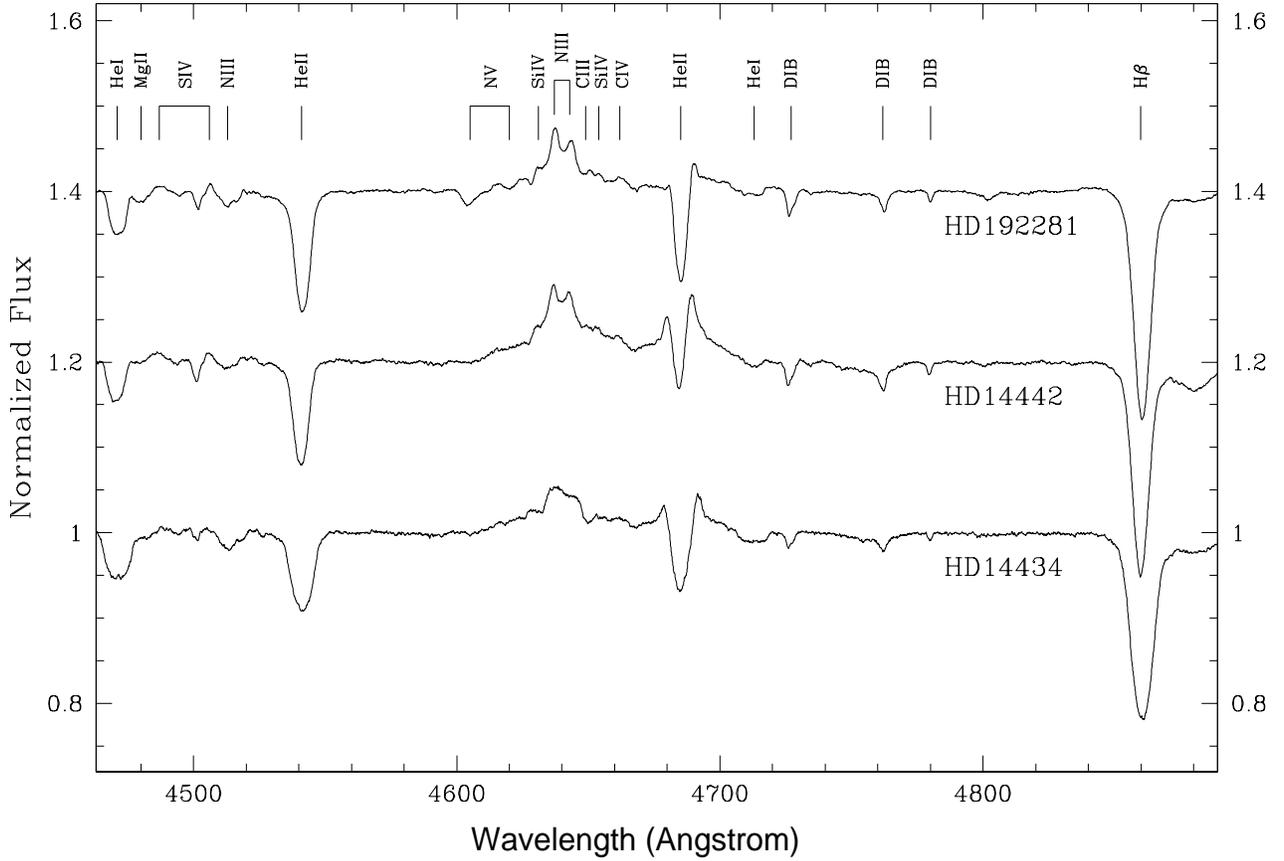}}
\end{center}
\caption{Mean spectra of our program stars between 4465 \AA\, and 4890 \AA\,. From top to bottom: HD\,192281: calculated from spectra of the September 2001 mission, between HJD 2\,452\,163.43 and HJD 2\,452\,170.41; HD\,14442: calculated from the spectra of the September 2002 mission, between HJD 2\,452\,518.54 and HJD 2\,452\,533.48; HD\,14434: calculated from the spectra of the October 2003 mission, between HJD 2\,452\,916.49 and HJD 2\,452\,934.38.\label{sp}}
\end{figure*}

\subsection{HD\,192281}
The mean blue spectrum of HD\,192281 is shown in Fig.\,\ref{sp}. The data of this spectrum are from the 2001 mission (see Table\,\ref{camp}). This part of the spectrum is dominated by absorptions of \ion{He}{i} $\lambda$ 4471, \ion{He}{ii} $\lambda$$\lambda$ 4541, 4686 and by the strong H$\beta$ line. \ion{Mg}{ii} $\lambda$ 4481 is clearly present, partially blended with \ion{He}{i} $\lambda$ 4471. In emission, the most striking feature is due to \ion{N}{iii} $\lambda\lambda$ 4634-41 and, to some extent, to \ion{C}{iii} $\lambda$ 4650 and probably \ion{C}{iv} $\lambda$ 4662. The two emission lines at $\lambda\lambda$ 4487,4506, identified as \ion{S}{iv} lines (Werner \& Rauch \cite{WR}), are also weakly present. We emphasize the probable presence of \ion{N}{v} absorption lines at 4605 and 4620 \AA\,. The presence of these lines in an O5 star is somewhat surprising. Indeed, these lines are commonly found in the spectra of very early (O3) stars (Walborn \& Fitzpatrick\,\cite{WF}). However, the lines are observed in the spectrum of HD14947 (O5If) as reported by Underhill et al. (\cite{UGH}), but not in the spectrum of HD15629 (O5V) discussed by the same authors. Some \ion{Si}{iv} emission lines at 4631 \AA\, and 4654 \AA\, are also present.\\
The double-peaked structure of the \ion{He}{ii} $\lambda$ 4686 emission justifies the spectral type given by Conti \& Leep (\cite{CL}). Such a line shape is supposed to result from a rotating wind (see e.g. Petrenz \& Puls \cite{PP}). Using the ratio of the equivalent widths (EWs) of the \ion{He}{i} $\lambda$ 4471 and \ion{He}{ii} 4541 lines ($\log{W'}$ = --0.47 $\pm$ 0.03) we assign an O5 spectral type, according to the classification criterion given by Mathys (\cite{Mat1}). In the following, we will adopt an O5(ef) spectral type for this star, where (ef) expresses the prominence of the central absorption component in the \ion{He}{ii} $\lambda$ 4686 line.

\subsection{HD\,14442}
Spectra of HD\,14442 have been taken during observing runs in July 1999, August 1999, September 2000, September 2001, September 2002 and October 2003. The middle spectrum in Fig.\,\ref{sp} shows the same general features as HD\,192281 except for the \ion{N}{v} lines. The central absorption of the \ion{He}{ii} $\lambda$ 4686 line appears less pronounced than in the case of HD\,192281. The strong emission wings observed for HD\,14442 confirm the Oef classification assigned by Conti \& Leep (\cite{CL}).\\
The equivalent width determined for the \ion{He}{i} $\lambda$ 4471 and \ion{He}{ii} 4541 lines allowed us to apply the $\log(W')$ criterion as defined by Mathys\,(\cite{Mat1}). The result (--0.41 $\pm$ 0.03) implies an O5.5 spectral type. In the following of this paper, HD\,14442 will be considered as an O5.5ef star.

\subsection{HD\,14434 \label{ms3}}
This star was observed during the September 2002 and October 2003 observing runs. The bottom spectrum of Fig.\,\ref{sp} displays the same general features as the spectrum of HD\,14442, but the lines are even further broadened by the fast stellar rotation. The fact that the \ion{N}{iii} $\lambda$$\lambda$ 4634 - 41 emission lines appear also significantly broader could imply that these lines are in fact formed at the level of the photosphere (rather than inside the stellar wind, see e.g. Mihalas \cite{Mih}). We further note that the fact that \ion{N}{iii} $\lambda$$\lambda$ 4634 - 41 lines appear in emission is not in agreement with the results of Kendall et al.\,(\cite{Ken2}) who reported the former weakly in absorption, as well as the absence of the latter. The double-peaked structure of the \ion{He}{ii} $\lambda$ 4686 line dominated by the central absorption, typical of O(ef), stars is obvious.\\
From the equivalent width of the \ion{He}{i} $\lambda$ 4471 and \ion{He}{ii} 4541 lines, we infer $\log(W')$ = --0.19 $\pm$ 0.04, corresponding to an O6.5 type. In the following of this study, we will thus adopt the O6.5(ef) spectral type for HD\,14434, in excellent agreement with the spectral type given by Conti \& Leep (\cite{CL}). 

\section{Radial velocities and equivalent widths \label{rvew}}
\subsection{HD\,192281}
\begin{figure}
\begin{center}
\resizebox{8.5cm}{4.5cm}{\includegraphics{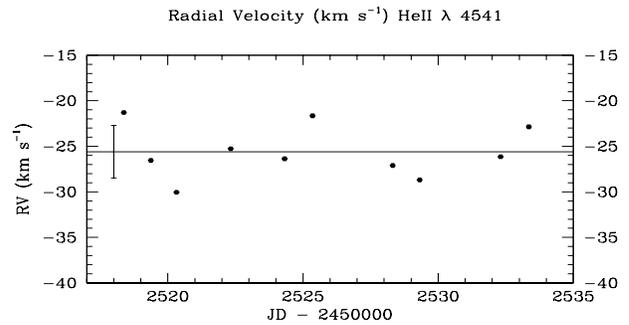}}
\end{center}
\caption{Radial velocities of HD\,192281 determined from the \ion{He}{ii} $\lambda$ 4541 line for the September 2002 observing run. The central wavelength was determined by fitting a gaussian to the line. The error bar gives an idea of the standard deviation obtained for the RV data set.\label{rv1}}
\end{figure}

As a first step, we have measured the radial velocity (RV) of the strongest absorption lines in the blue range. The lines of interest here are \ion{He}{ii} $\lambda$ 4541 and H$\beta$. Since the \ion{He}{ii} $\lambda$ 4541 line forms probably deep inside the photosphere, it is likely to provide the most reliable indicator of the actual stellar RV. As shown in Fig.\,\ref{rv1}, the dispersion in radial velocity reaches a maximum of about 10 km\,s$^{-1}$, whereas the typical error\footnote{The error on the radial velocity estimate corresponds to the standard deviation determined for the radial velocity of a Diffuse Interstellar Band (DIB) at about 4762 \AA\,.} in the determination of RVs with our gaussian fitting method is about 8 km\,s$^{-1}$. For all our data sets, we find a mean radial velocity of $-$23.1 km\,s$^{-1}$. This leads to a value of about 5.0 km\,s$^{-1}$ for the peculiar velocity (after application of the LSR and Galactic rotation corrections). This value strongly argues against a runaway status for this star. Indeed, according to the criteria of Stone (\cite{Sto}), the peculiar velocity threshold for a star to be a runaway is about 40 km\,s$^{-1}$. It has been suggested by Gies \& Bolton (\cite{GB}) that the runaway status was erroneously attributed to stars such as HD\,192281 because of their huge wind velocity that yield a large negative RV for those spectral lines that are at least partly formed in the wind. The choice of the line for RV estimation is thus particularly crucial.\\
Gies \& Bolton (\cite{GB}) measured also the RV of the \ion{He}{ii} $\lambda$ 4686 and hydrogen lines for this star. After correction for the velocity component due to the solar motion in the LSR frame, they found a mean of $-$26.8 km\,s$^{-1}$ for the \ion{He}{ii} $\lambda$ 4686 line, and a mean of $-$44.7 km\,s$^{-1}$ for hydrogen lines. For comparison, we find LSR-corrected RV means of $-$33.3 and $-$46.1 km\,s$^{-1}$ respectively for \ion{He}{ii} $\lambda$ 4686 and H$\beta$. The reasonable agreement between our values and those of Gies \& Bolton suggests that there are no large long-term variations of the RVs. The difference of about 18 km\,s$^{-1}$ between the mean RV of these two lines (about 13 km\,s$^{-1}$ in our case) is again explained by a difference in the depth of the line formation region within the expanding atmosphere. Indeed, the \ion{He}{ii} $\lambda$ 4686 line is most probably formed deeper in the wind, where the temperature is higher, and the wind velocity is lower.

\begin{table*}
\caption{Radial velocities and equivalent widths of the strongest absorption lines in the spectrum of HD\,192281. The RV of the \ion{He}{i} $\lambda$ 4471 line is not quoted because its shape deviates significantly from a Gaussian, leading to a poor estimate of its central position. The complex shape of the \ion{He}{ii} $\lambda$ 4686 line, and its strong variability, prevented us to derive any useful EW for this line. The error bars represent the $\pm$\,1\,$\sigma$ standard deviation. An estimate of the mean signal-to-noise ratio of an individual spectrum is given in the last column.\label{rvew1}}
\begin{center}
\begin{tabular}{l c c c c c c c}
\hline
\hline
 Mission & \ion{He}{i} $\lambda$ 4471 & \multicolumn{2}{c}{\ion{He}{ii} $\lambda$ 4541} & \ion{He}{ii} $\lambda$ 4686 & \multicolumn{2}{c}{H$\beta$} & S/N\\
 & EW & RV & EW & RV & RV & EW & \\
 & (\AA) & (km\,s$^{-1}$) & (\AA) & (km\,s$^{-1}$) & (km\,s$^{-1}$) & (\AA) & \\
\hline
September 1998 & 0.31 $\pm$ 0.04 & --22.4 $\pm$ 3.0 & 0.84 $\pm$ 0.04 & --48.9 $\pm$ 7.4 & $-$ & $-$ & 380 \\
Summer 1999 & 0.28 $\pm$ 0.01 & --21.5 $\pm$ 2.4 & 0.81 $\pm$ 0.02 & --43.7 $\pm$ 9.2 & --62.8 $\pm$ 3.5 & 1.97 $\pm$ 0.03 & 290 \\
September 2001 & 0.31 $\pm$ 0.01 & --22.9 $\pm$ 2.8 & 0.92 $\pm$ 0.01 & --44.3 $\pm$ 5.0 & --56.5 $\pm$ 2.4 & 2.05 $\pm$ 0.03 & 450 \\
September 2002 & 0.30 $\pm$ 0.01 & --25.6 $\pm$ 2.9 & 0.94  $\pm$ 0.01 & --48.6 $\pm$ 6.0 & --58.1 $\pm$ 3.9 & 2.05 $\pm$ 0.02 & 440 \\
October 2003 & 0.31 $\pm$ 0.01 & --13.8 $\pm$ 5.2 & 0.88 $\pm$ 0.02 & --41.8 $\pm$ 12.8 & --58.1 $\pm$ 3.2 & 2.06 $\pm$ 0.04 & 380 \\
\hline
\end{tabular}
\end{center}
\end{table*}

\begin{table*}
\caption{Same as Table\,\ref{rvew1}, but for HD\,14442.\label{rvew2}}
\begin{center}
\begin{tabular}{l c c c c c c c}
\hline
\hline
 Mission & \ion{He}{i} $\lambda$ 4471 & \multicolumn{2}{c}{\ion{He}{ii} $\lambda$ 4541} & \ion{He}{ii} $\lambda$ 4686 & \multicolumn{2}{c}{H$\beta$} & S/N\\
 & EW & RV & EW & RV & RV & EW & \\
 & (\AA) & (km\,s$^{-1}$) & (\AA) & (km\,s$^{-1}$) & (km\,s$^{-1}$) & (\AA) & \\
\hline
Summer 1999 & 0.33 $\pm$ 0.02 & --30.3 $\pm$ 8.2 & 0.84 $\pm$ 0.02 & --51.5 $\pm$ 18.4 & --67.7 $\pm$ 3.7 & 1.84 $\pm$ 0.04 & 160 \\
September 2000 & 0.32 $\pm$ 0.02 & --32.2 $\pm$ 6.7 & 0.84 $\pm$ 0.02 & --59.7 $\pm$ 14.9 & --66.6 $\pm$ 5.1 & 1.91 $\pm$ 0.03 & 340 \\
September 2001 & 0.34 $\pm$ 0.01 & --31.8 $\pm$ 5.6 & 0.84 $\pm$ 0.01 & --57.1 $\pm$ 23.7 & --64.5 $\pm$ 3.8 & 1.86 $\pm$ 0.03 & 300 \\
September 2002 & 0.32 $\pm$ 0.01 & --34.3 $\pm$ 2.9 & 0.83  $\pm$ 0.02 & --66.5 $\pm$ 16.2 & --65.7 $\pm$ 4.3 & 1.84 $\pm$ 0.05 & 430 \\
October 2003 & 0.32 $\pm$ 0.01 & --32.0 $\pm$ 10.7 & 0.80 $\pm$ 0.01 & --57.4 $\pm$ 19.6 & --64.5 $\pm$ 14.5 & 1.84 $\pm$ 0.05 & 330 \\
\hline
\end{tabular}
\end{center}
\end{table*}

\begin{table*}
\caption{Same as Table\,\ref{rvew1}, but for HD\,14434.\label{rvew3}}
\begin{center}
\begin{tabular}{l c c c c c c c}
\hline
\hline
 Mission & \ion{He}{i} $\lambda$ 4471 & \multicolumn{2}{c}{\ion{He}{ii} $\lambda$ 4541} & \ion{He}{ii} $\lambda$ 4686 & \multicolumn{2}{c}{H$\beta$} & S/N\\
 & EW & RV & EW & RV & RV & EW & \\
 & (\AA) & (km\,s$^{-1}$) & (\AA) & (km\,s$^{-1}$) & (km\,s$^{-1}$) & (\AA) & \\
\hline
September 2002 & 0.54 $\pm$ 0.03 & --15.3 $\pm$ 5.7 &  0.81 $\pm$ 0.03 & --40.2 $\pm$ 18.2 & --39.4 $\pm$ 3.5 & 2.09 $\pm$ 0.03 & 530 \\
October 2003 & 0.52 $\pm$ 0.02 & --3.9 $\pm$ 12.0 & 0.84 $\pm$ 0.03 & --28.4 $\pm$ 13.4 & --35.3 $\pm$ 5.3 & 2.10 $\pm$ 0.05 & 320 \\
\hline
\end{tabular}
\end{center}
\end{table*}

Because of its profile variability, which will be discussed hereafter, the \ion{He}{ii} $\lambda$ 4686 line yields less reliable RVs. Indeed, the radial velocities spread over more than 35 km\,s$^{-1}$. We did however not find any systematic trend for this variability in radial velocities.\\
Barannikov (\cite{Bar}) performed a RV variability analysis, the only known, for this star. He claimed to find RV variation with a semi-amplitude of 16.1 $\pm$ 2.4 km\,s$^{-1}$ on a 5.48 days period, compatible with the existence of a low mass companion. We have folded our RV data with Barannikov's 5.48 days period but no systematic trend appeared. A Fourier analysis was also performed on our RV measurements of the \ion{He}{ii} $\lambda$ 4541 line. The resulting periodogram does not show any peak with an amplitude larger than 2.4 km\,s$^{-1}$. The highest peak appears at a frequency of 0.105 d$^{-1}$, corresponding to a 9.57 d period. This value is quite close to the period (9.59 d) obtained by Barannikov (\cite{Bar}) for his photometric data. In fact, Barannikov reported also apparently periodic light variations with a peak to peak amplitude of about 0.04 mag. This coincidence is somewhat puzzling. Although this could be regarded as evidence for binarity, we do not consider this to be a plausible scenario. In fact, the amplitude of the RV variation ($\sim$2.5 km\,s$^{-1}$) at the 9.57 d `period' is of the order of the uncertainty on individual data points. Moreover, it seems extremely difficult to account simultaneously for a light curve of amplitude $\sim$0.04 mag (peak to peak) and a RV variation inferior to about 5 km\,s$^{-1}$ (peak to peak) within a binary scenario. Assuming that the light variations are due to total eclipses of a companion, we could be dealing with a secondary of spectral type $\sim$\,B3V and such an object would produce a RV curve with a peak to peak amplitude of $\geq$\,80 km\,s$^{-1}$ (assuming i $\geq$ 60$^\circ$). On the other hand, to account for the amplitude of the RV curve, we would need a companion of mass inferior to about 1 M$_\odot$ and such a star would be far too faint to produce a variation of 0.04 mag in the light curve, even if it were still on a pre-main sequence track.\\ 
The equivalent width was measured for the \ion{He}{i} $\lambda$ 4471, \ion{He}{ii} $\lambda$ 4541 and H$\beta$ lines. The results, as well as the RVs, are collected in Table\,\ref{rvew1}. Mean values are given for each observing run with an estimate of the standard deviation. We see that no significant variability appears from one run to the other. Moreover, no significant variability was found during individual observing campaigns.

\subsection{HD\,14442}
\begin{figure}
\begin{center}
\resizebox{8.5cm}{4.5cm}{\includegraphics{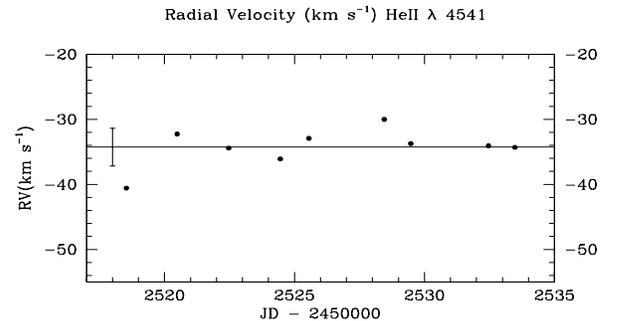}}
\end{center}
\caption{Same as Fig.\,\ref{rv1}, but for HD\,14442.\label{rv2}}
\end{figure}

Gaussians were fitted to the \ion{He}{ii} $\lambda$ 4541 line to determine the radial velocity. The results are illustrated in Fig.\,\ref{rv2} which shows the RV versus heliocentric Julian day for the September 2002 observing run. We find, for this line, a mean RV of about $-$32.1 km\,s$^{-1}$ for the whole data set, which is somewhat different from the values reported by other authors such as Conti et al. (\cite{CLL}) who give $-$42.1 km\,s$^{-1}$ or Underhill \& Gilroy (\cite{UG}) who published a value of $-$40.6 km\,s$^{-1}$, for averages of several line measurements. As for the differences in RVs found for different lines, the discrepancy between our results and those of the authors cited hereabove can probably be explained by the velocity gradient in the stellar wind. The maximum difference in RV over the five data sets is only about 30 km\,s$^{-1}$. A Fourier analysis was performed on these RV data but no peak with an amplitude larger than 4.4 km\,s$^{-1}$ emerged from the periodogram.\\
RVs were also determined for the \ion{He}{ii} $\lambda$ 4686 and H$\beta$ lines. For the first one, RV values vary around a mean of $-$57.0 km\,s$^{-1}$, whilst for H$\beta$, this mean is about $-$66.3 km\,s$^{-1}$. As for HD\,192281, the mean RV is more negative in the case of the hydrogen line.

The equivalent width was determined for the same lines as for HD\,192281. Once again, no significant variability is found during the runs, or from one observing run to the other.

\subsection{HD\,14434}
\begin{figure}
\begin{center}
\resizebox{8.5cm}{4.5cm}{\includegraphics{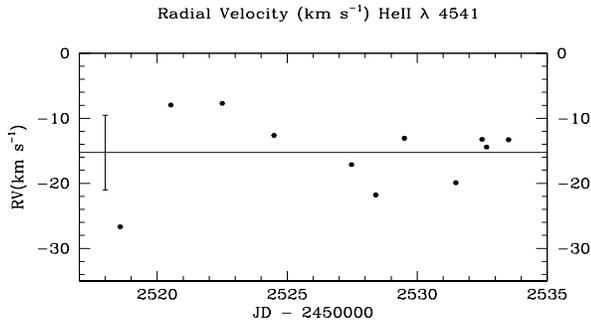}}
\end{center}
\caption{Same as Fig.\,\ref{rv1}, but for HD\,14434.\label{rv3}}
\end{figure}

Radial velocities measured on the same lines as for HD\,192281 and HD\,14442 are given in Table\,\ref{rvew3}. We emphasize that these values are subject to larger uncertainties due the difficulty to fit gaussians to the very broad absorption of this star: the actual line profiles are not gaussian and undergo variations which are discussed in more details in Sect.\,\ref{var3}. The rather large values of the standard deviations given in Table\,\ref{rvew3} reflect these difficulties. The largest error bars are obtained for the \ion{He}{ii} $\lambda$ 4686 line, and are undoubtedly due to the strong variability of this line.\\
Fig.\,\ref{rv3} shows the radial velocities measured on the \ion{He}{ii} $\lambda$ 4541 line during the September 2002 observing run as a function of time. The dispersion is rather large, but no systematic trend can be detected. Mean radial velocities for the September 2002 and October 2003 campaigns overlap within their large error bars. The radial velocity reported by Conti et al. (\cite{CLL}), i.e. --11.7 km\,s$^{-1}$, is compatible with the value obtained for the \ion{He}{ii} $\lambda$ 4541 line. No indication of binarity on time scales of several days, or from one year to the next, has thus been revealed by our time series.\\
Equivalent width are also quoted in Table\,\ref{rvew3}. These EWs are stable on the time scales investigated in our study. 

\section{Line profile variability \label{lpv}}
The first step of our line variability analysis (see also Rauw et al.\,\cite{RDV}) is the identification of the lines which vary significantly with time. For this purpose, we used the time variance spectrum (TVS) as developed by Fullerton et al.\,(\cite{Ful}). We evaluated the signal-to-noise ratio of individual spectra over a line free region between 4560 \AA\, and 4590 \AA\,. This estimate is necessary for the determination of the 99 \% significance level of the variability found with the TVS.\\
The procedure was applied to each dataset. The lines that show significant (at the 99 \% confidence level) variability are essentially \ion{He}{ii} $\lambda$ 4686 and H$\beta$. The detailed line profile variability of these two lines is described hereafter.\\

In order to quantify the variability of the various spectral lines, we performed a 2D Fourier analysis using the technique described by Rauw et al. (\cite{Rauw}): at each wavelength step, the detection of candidate frequencies is performed by an unequally spaced data Fourier analysis as described by Heck et al.\,(\cite{HMM}). The main peaks ($\nu$) and their aliases (essentially 1$-\nu$) are identified and are considered for further analysis. As a next step we `prewhite' the data for the variations at frequency $\nu$ following a method already described by Rauw et al.\,(\cite{Rauw}) and used in Paper I. The frequencies which give the best results, i.e. a periodogram devoid of significant peaks after prewhitening, are considered as good candidates. Ideally, a physically significant frequency should meet the following criteria:
\begin{enumerate}
\item A stable clock ruling the variability should be found in the data of a given line from more than one observing campaign.
\item The frequency should not be found in the periodogram of a constant line (typically an interstellar feature), otherwise such a frequency probably reflects an artificial variability affecting the entire spectrum, and should not be associated with a physical process. 
\item The frequency should appear, within the error bars, in the periodograms of more than one variable line if more than one line varies significantly with time.
\end{enumerate}
These criteria concern in fact only the most ideal cases and have to be relaxed in more complex situations (like the case of BD\,+60$^\circ$ 2522, Paper I). Their strict application would probably lead to the rejection of potentially meaningful frequencies. The last criterion, in particular, is certainly not mandatory because we cannot state that a given variability mechanism will affect all lines in a same manner.

\subsection{HD\,192281}
An overview of the observing runs is given in Table\,\ref{runs1}. Generally, one spectrum was obtained during each night the star was observed, except for the 2001 run where at least 2 spectra were taken each night with a mean time separation of about 0.025 d.
\begin{table}
\caption{Observing runs used for the line profile variability study of HD\,192281. The first column gives the name of the campaign as used in the text, the second column yields the amount of spectra taken during each run, the third specifies which line was studied with the corresponding data set, the fourth one gives the time elapsed between the first and the last spectrum of the run, and finally the last column provides the natural width of a peak of the power spectrum taken as 1/$\Delta$T.\label{runs1}}
\begin{center}
\begin{tabular}{l c c c c}
\hline
\hline
 Mission & Number of & Line & $\Delta$T & $\Delta\nu_\mathrm{nat}$\\
 & spectra & & (d) & (d$^{-1}$)\\
\hline
Sept. 1998 & 6 & \ion{He}{ii} $\lambda$ 4686 & 7.977 & 0.125\\
Sum. 1999 & 11 & \ion{He}{ii} $\lambda$ 4686 & 33.140 & 0.030\\
	& & H$\beta$ & & \\
Sept. 2001 & 16 & \ion{He}{ii} $\lambda$ 4686 & 6.979 & 0.143\\
	& & H$\beta$ & & \\
Sept. 2002 & 10 & \ion{He}{ii} $\lambda$ 4686 & 14.984 & 0.067\\
	& & H$\beta$ & & \\
Oct. 2003 & 7 & \ion{He}{ii} $\lambda$ 4686 & 15.910 & 0.063\\
	& & H$\beta$ & & \\
\hline
\end{tabular}
\end{center}
\end{table}

\begin{figure}
\begin{center}
\resizebox{8.5cm}{8.0cm}{\includegraphics{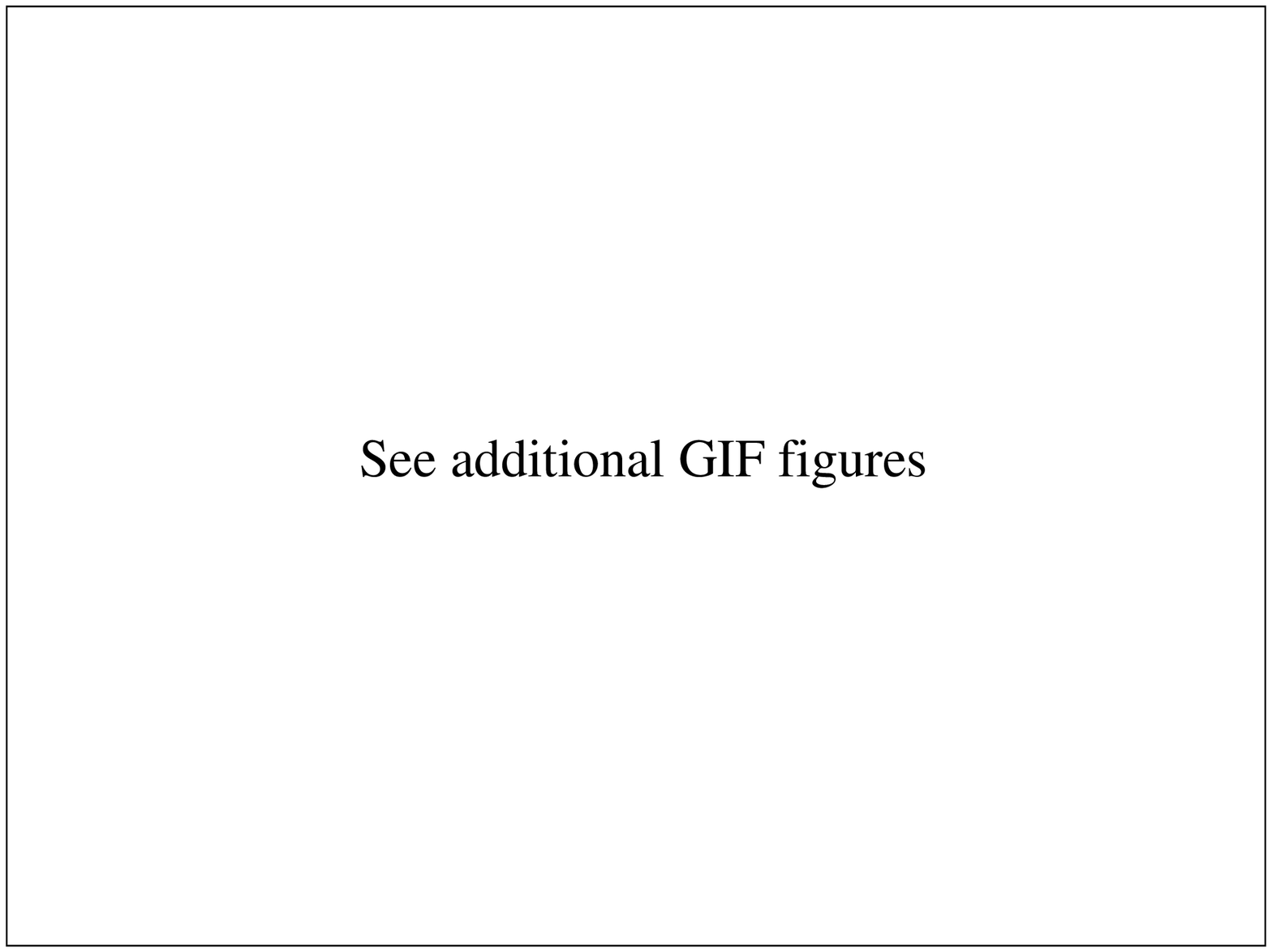}}
\end{center}
\caption{Upper panels: mean spectrum and square root of the time variance spectrum of the \ion{He}{ii} $\lambda$ 4686 line of HD\,192281 between 4675 \AA\, and 4695 \AA\, respectively for the September 2001 observing run and for the combined data set. The 99 \% significance level of the TVS$^{1/2}$ is indicated by the horizontal dotted line. Velocities are expressed in km\,s$^{-1}$. Lower panels: the corresponding two-dimensional power spectra.  The darkest regions stand for the highest peak intensities of the periodogram. Frequencies and wavelengths are expressed respectively in d$^{-1}$ and in \AA\, (see figure 0548f5.gif). \label{F2D1}}
\end{figure}

\paragraph{\ion{He}{ii} $\lambda$ 4686.}
As shown in Table\,\ref{runs1}, five data sets were available for the study of this line. TVS and Fourier analyses were performed on the five time series, as well as on a combination of all our data sets. For a better visualization of the behaviour of this line, Fig.\,\ref{F2D1} shows the mean spectrum and the TVS, as well as the 2-D power spectrum as a grey scale representation for the September 2001 observing run, and in the case of the combined data set. The periodogram of the combined data set displays some kind of fine structure due to the strong irregularity of the sampling. The TVS calculated for all our individual data sets reveals that the level of the variability changes from one observing run to the other. The variability affects essentially two separate parts of the line, respectively the violet and the red wings.

\begin{figure}
\begin{center}
\resizebox{8.5cm}{10.0cm}{\includegraphics{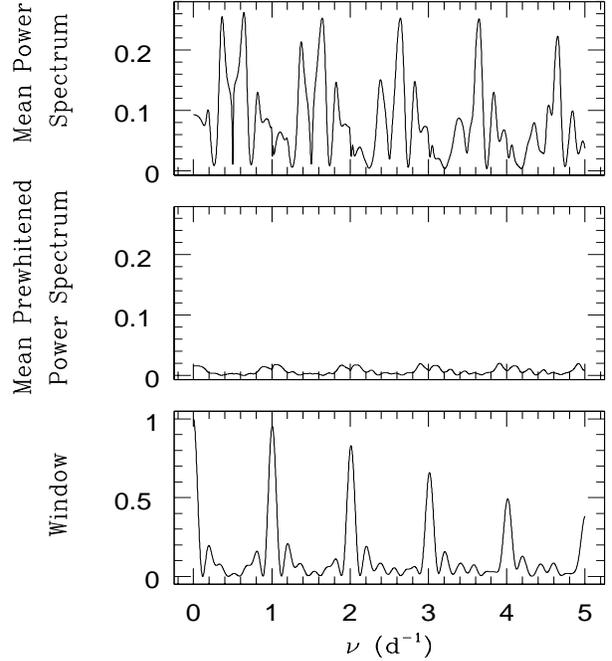}}
\end{center}
\caption{Upper panel: mean power spectrum obtained for the \ion{He}{ii} $\lambda$ 4686 line of HD\,192281 between 4687.5 \AA\, and 4688.5 \AA\, for the September 2001 observing run, up to 5.0 d$^{-1}$. Middle panel: residual power spectrum obtained after prewhitening with a frequency of 0.64 d$^{-1}$. Bottom panel: power spectral window showing secondary peaks which explain the asymmetrical shape of the peaks of the power spectrum.\label{prew1}}
\end{figure}

The most striking and recurrent feature in these periodograms is the presence of aliases at frequencies of about 0.35 and 0.65 d$^{-1}$, corresponding respectively to periods of about 2.86 and 1.54 d. Figure\,\ref{prew1} shows the mean periodogram obtained for the 2001 run between 0.0 and 5.0 d$^{-1}$, along with the residual periodogram obtained after prewhitening for a frequency of 0.64 d$^{-1}$ between 4687.5 and 4688.5 \AA\,. The spectral window is displayed in the bottom panel. We emphasize that the time analysis performed on such a narrow wavelength band does not provide any result representative of the behaviour of the whole line, but it allows a clear identification of the time scale of the variability observed in the part of the line where the TVS reaches its highest level. Prewhitening shows that most of the power of the periodogram can be accounted for by aliasing of this 0.64 d$^{-1}$ frequency (or its $1-\nu$ alias).

The periodogram of a Diffuse Interstellar Band (DIB) at 4762 \AA\, has been calculated for each data set, and only the September 1998 campaign gives peaks situated near these candidate frequencies, but with much lower intensities.\\
The variability pattern of the \ion{He}{ii} $\lambda$ 4686 line changes with epoch. Indeed, the strongest variability and hence the $\nu_1$ and 1$-\nu_1$ peaks appear essentially in the red and the violet (with lower intensity) wings in the September 1998 and September 2001 periodograms, but appear only in the violet wing in Summer 1999. The width of the wavelength range displaying a significant variability with frequency $\nu_1$ or 1$-\nu_1$ changes also from one epoch to the other. A summary of all the frequencies is given in Table\,\ref{rec} (Sect.\,\ref{disc}).

\paragraph{H$\beta$.} 
The same approach as for the \ion{He}{ii} $\lambda$ 4686 line was adopted for H$\beta$. Our data allow us to perform this line profile variability analysis only for the 1999, 2001, 2002 and 2003 campaigns. In this case, the TVS and the 2-D power spectra reveal that the variability is concentrated in the core of the line. The Summer 1999 data set analysis leads once again to frequencies of about 0.33 or 0.67 d$^{-1}$. This observation raises thus the degree of confidence concerning these frequencies already noted for the \ion{He}{ii} $\lambda$ 4686 line. However, the situation is less clear for the September 2001 periodogram. Indeed, the peak frequencies found in this case are near 0.14 and 0.46 d$^{-1}$. For the September 2002 campaign, the variability appears shifted to the red wing of the line, and we are not able to definitely distinguish the peaks from those which appear in the DIB periodogram.

\paragraph{\ion{He}{i} $\lambda$ 4713.}
Even if no variability is seen for this line until 2001, the September 2002 observing campaign reveals a significant TVS signal in the red wing. The existence of this variability turns out to be robust in the sense that it remains if we omit individual spectra one at the time and repeat the analysis on the remaining data set. Although the periodogram is dominated by a peak at $\sim$0.3 d$^{-1}$, prewhitening at this frequency yields no satisfactory result.

\subsection{HD\,14442}
Table\,\ref{runs2} gives some information about the observing runs. During the September 2001 mission, two spectra were taken each night with a mean separation between the two spectra of about 0.025 d.
\begin{table}
\caption{Same as Table\,\ref{runs1}, but for the observing runs used for the line profile variability study of HD\,14442. \label{runs2}}
\begin{center}
\begin{tabular}{l c c c c}
\hline
\hline
 Mission & Number of & Line & $\Delta$T & $\Delta\nu_\mathrm{nat}$ \\
 & spectra & & (d) & (d$^{-1}$) \\
\hline
Sum. 1999 & 11 & \ion{He}{ii} $\lambda$ 4686 & 34.039 & 0.029 \\
	& & H$\beta$ & & \\
Sept. 2000 & 9 & \ion{He}{ii} $\lambda$ 4686 & 10.967 & 0.091 \\
	& & H$\beta$ & & \\
Sept. 2001 & 12 & \ion{He}{ii} $\lambda$ 4686 & 7.030 & 0.142 \\
	& & H$\beta$ & & \\
Sept. 2002 & 9 & \ion{He}{ii} $\lambda$ 4686 & 14.942 & 0.067 \\
	& & H$\beta$ & & \\
Oct. 2003 & 6 & \ion{He}{ii} $\lambda$ 4686 & 17.957 & 0.056 \\
	& & H$\beta$ & & \\
\hline
\end{tabular}
\end{center}
\end{table} 

\begin{figure}
\begin{center}
\resizebox{8.5cm}{8.0cm}{\includegraphics{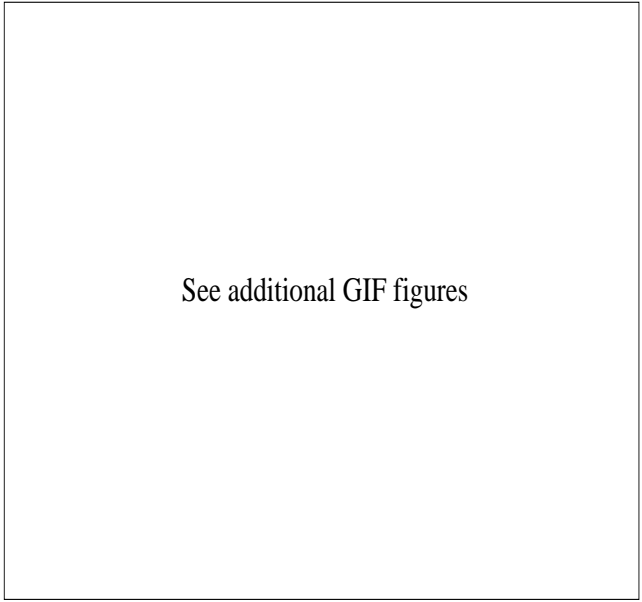}}
\end{center}
\caption{Same as Figure\,\ref{F2D1} but for HD\,14442 (see figure 0548f8.gif).\label{F2D2}}
\end{figure}

\paragraph{\ion{He}{ii} $\lambda$ 4686.}
In Fig.\,\ref{F2D2}, we show the results obtained for the September 2001 campaign, as well as for the combined data set. The analysis of the individual data sets reveals that the TVS obtained for the September 2002 campaign shows a more complex structure, with an additional varying component in the blue emission peak. To check the robustness of this feature, we have removed each individual spectrum one at the time from our September 2002 time series and performed the variability analysis on the remaining data. This additional component in the TVS appears to be stable. The situation is even more complex in the case of the October 2003 data set, where we can distinguish four distinct varying parts in the line, but it could be due to a rather poor sampling. Due to the poor weather conditions we obtained indeed only 6 spectra during that campaign.\\
All our data sets (except perhaps in Summer 1999 where the variability is much less significant) yield a relatively stable frequency between 0.35 and 0.40 d$^{-1}$ (or 0.65 and 0.60 d$^{-1}$ for their respective aliases) near the line core. These frequencies give very good prewhitening results and are not the same as those of the main peaks appearing for the DIB at 4762 \AA\,.

\begin{figure}
\begin{center}
\resizebox{8.5cm}{10.0cm}{\includegraphics{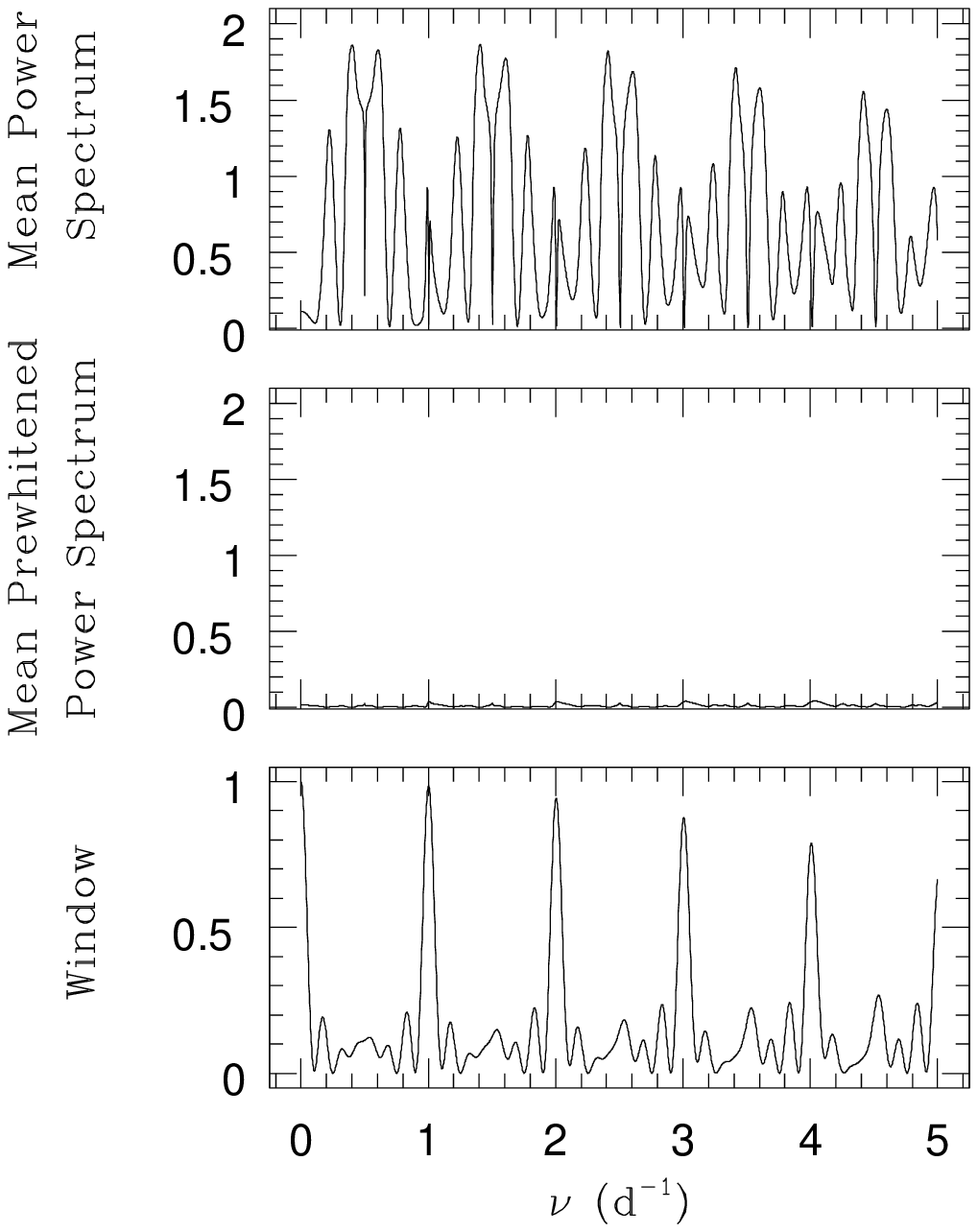}}
\end{center}
\caption{Upper panel: mean power spectrum obtained for the \ion{He}{ii} $\lambda$ 4686 line of HD\,14442 between 4682 \AA\, and 4684 \AA\, for the September 2001 observing run, up to 5.0 d$^{-1}$. Middle panel: residual power spectrum obtained after prewhitening with a frequency of 0.60 d$^{-1}$. Bottom panel: power spectral window showing secondary peaks which explain the asymetrical shape of the peaks of the power spectrum.\label{prew2}}
\end{figure}

\begin{figure}
\begin{center}
\resizebox{8.5cm}{8.0cm}{\includegraphics{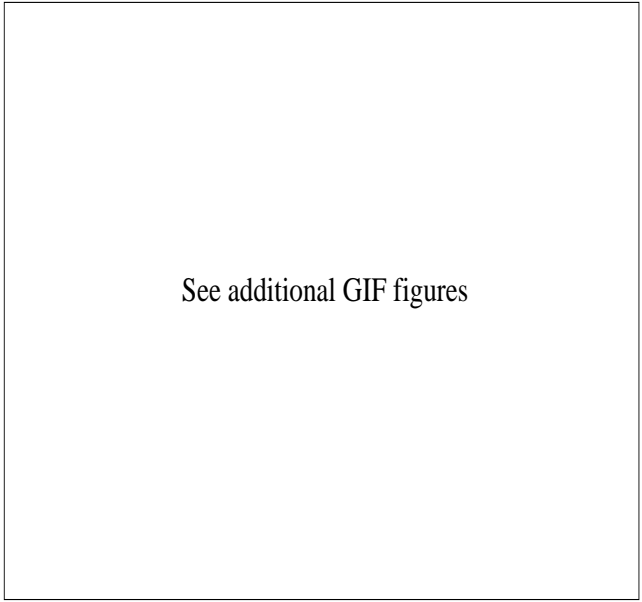}}
\end{center}
\caption{Same as Figure\,\ref{F2D2} but for the H$\beta$ line between 4845 \AA\, and 4875 \AA\, (see figure 0548f9.gif).\label{F2D2b}}
\end{figure}

Fig.\ref{prew2} shows the mean power spectrum obtained for the September 2001 mission between 4682 \AA\, and 4684 \AA\,. The middle panel of this figure illustrates the residual periodogram obtained after prewhitening with a frequency of 0.60 d$^{-1}$. It appears that the aliasing of this frequency is sufficient to account for all the peaks of the power spectrum. A similar result is achieved if the prewhitening is performed with a 0.40 d$^{-1}$ frequency.\\

Other peaks appear also in our periodograms. For example, a frequency of 0.11 d$^{-1}$ (and its aliases) is clearly identified in the red wing for the Summer 1999 campaign. Table\,\ref{rec} in Sect.\,\ref{disc} gives a more complete overview of the frequencies emerging from the analysis of all our time series.

\paragraph{H$\beta$.}
No significant variability was found for this line in the Summer 1999 data. Remember that this data set has the poorer signal-to-noise ratio, and also the poorer sampling due to the spread of our observation over 34 days. However, significant variations were found for the other observing runs, especially for the September 2002 and the October 2003 campaigns where it reaches its highest level. The two-dimensional periodograms in Fig.\,\ref{F2D2b} show variability patterns which are very similar to those obtained for the \ion{He}{ii} $\lambda$ 4686 line. The simplest case appears for the 2001 data set, where the main feature is completely reproduced by a $~$0.39 d$^{-1}$ (or 0.60 d$^{-1}$) frequency. This is very interesting if we remember that it corresponds quite well to the frequencies obtained for the \ion{He}{ii} $\lambda$ 4686 line. The situation is more complex in the case of the 2000 and 2002 campaigns. After analysis, it appears that the periodograms are partly explained by some complex long term behaviour probably due to the sampling or normalization irregularities. Nevertheless, frequencies of about 0.35 d$^{-1}$ (or 0.65 d$^{-1}$) obviously contribute to the power spectra obtained in both cases. The October 2003 periodogram is mainly dominated by peaks at about 0.41 d$^{-1}$ (0.59 d$^{-1}$).

\paragraph{\ion{He}{ii} $\lambda$ 4541.}
In the 2002 campaign, the good quality of the data allowed us to find a significant (albeit weak) line profile variability in the case of the \ion{He}{ii} $\lambda$ 4541 line, especially near the core of the line with some weak contributions in the red wing. The analysis of the periodogram revealed that frequencies between 0.36 and 0.4 d$^{-1}$ (or their 1$-\nu$ aliases) contribute significantly to the variability of this line. This range of frequencies matches very well the frequencies obtained for the \ion{He}{ii} $\lambda$ 4686 and the H$\beta$ lines discussed hereabove. This result reinforces our trust in the physical origin of the frequencies found for the variations in the spectrum of HD\,14442.

\subsection{HD\,14434 \label{var3}}

HD\,14434 was observed during only two observing runs (see Table\,\ref{runs3}). In order to search for rapid variations, the star was intensively observed during the night from October 16 to 17, 2003: ten spectra spanning 6 hours were obtained. The TVS for this specific night computed over the whole spectrum did not reveal any significant variability. Consequently, we conclude that HD\,14434 does not display a variable behaviour on short time scales (a few hours) in the spectral range we investigated.\\
However, over time scales of several days, a significant variability is detected essentially for the \ion{He}{ii} $\lambda$ 4686 line in both data sets, and for H$\beta$ in October 2003. The behaviour of these two lines was investigated and the results are described hereafter. The high quality of our September 2002 data also allowed us to detect some variability for the \ion{He}{i} $\lambda$ 4471 line, but with a level which is too low for a Fourier analysis.

\begin{figure}
\begin{center}
\resizebox{8.5cm}{8.0cm}{\includegraphics{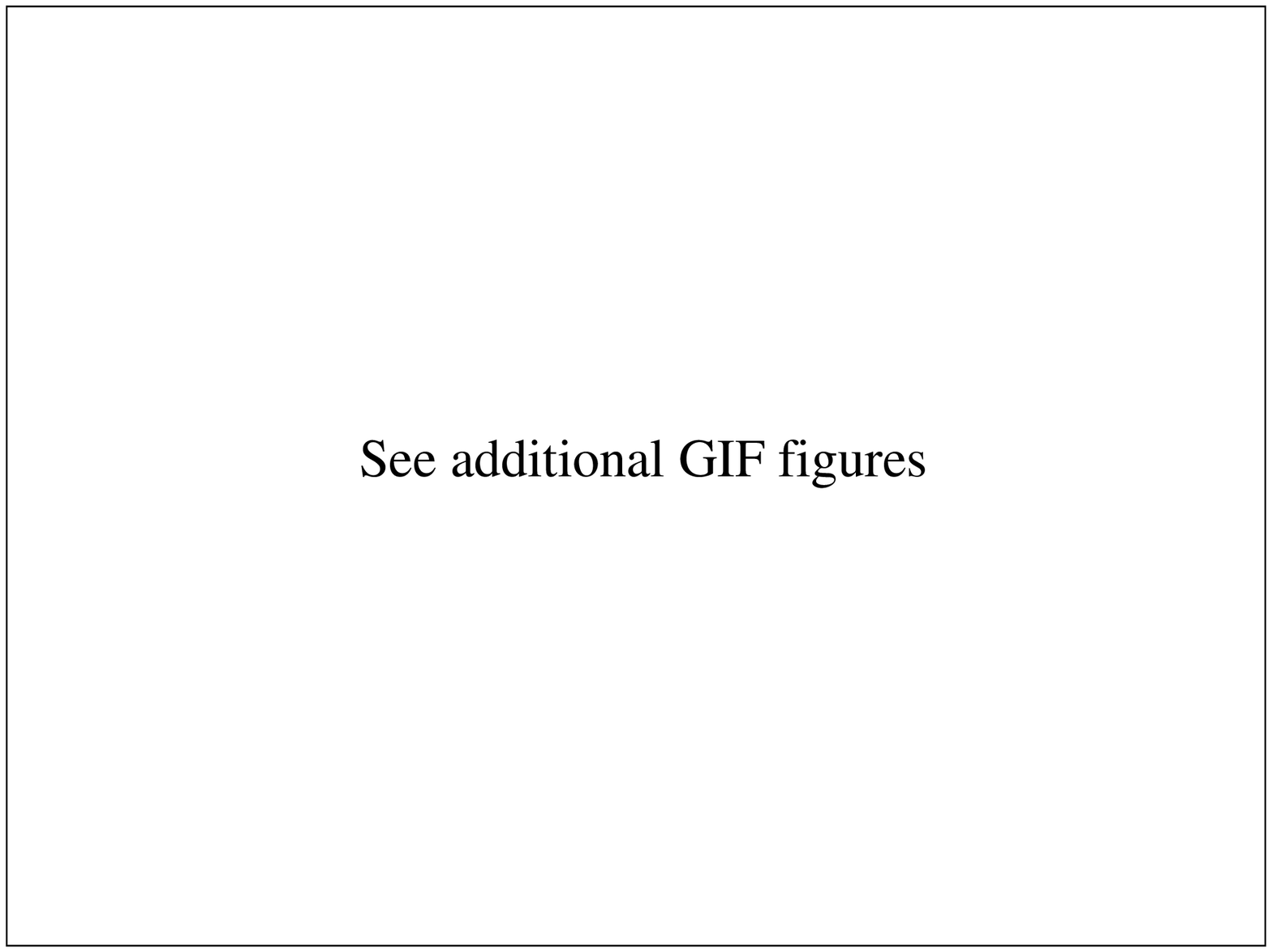}}
\end{center}
\caption{Upper panels: mean spectrum and square root of the time variance spectrum of the \ion{He}{ii} $\lambda$ 4686 line of HD\,14434 between 4675 \AA\, and 4695 \AA\, respectively for the September 2002 and the October 2003 observing runs. The 99 \% significance level of the TVS$^{1/2}$ is indicated by the horizontal dotted line. The differences in significance levels are due to differences in signal-to-noise ratios (see Table\,\ref{rvew3}). Velocities are expressed in km\,s$^{-1}$. Lower panels: the corresponding two-dimensional power spectrum.  The darkest regions stand for the highest peak intensities of the periodogram. Frequencies and wavelengths are expressed respectively in d$^{-1}$ and in \AA\, (see figure 0548f10.gif). \label{F2D3}}
\end{figure}

\paragraph{\ion{He}{ii} $\lambda$ 4686.}
In Fig.\,\ref{F2D3}, the TVS shows that a significant variability is detected for the \ion{He}{ii} $\lambda$ 4686 line in both data sets. The variability level seems higher in October 2003 than in September 2002. The variations are not restricted to a given part of the line, but extend over a rather broad domain. Lower panels of the same figure display the two-dimensional power spectra obtained through Fourier analysis. The behaviour of the line is rather complex, and the periodograms reveal that different parts of the line apparently display different variability trends. A more detailed analysis, i.e. by focusing on individual parts of the line, yields different time scales.\\
The September 2002 power spectrum reveals that the dominant frequency is about 0.91-0.92 d$^{-1}$. This value corresponds to the darkest spots of the lower left panel of Fig.\,\ref{F2D3}. We note that the validity of the corresponding periods (close to one day) can not be totally warranted by our sampling. Other peaks contribute also to the power but the prewhitening failed to reveal unambigously which ones were to be selected to identify the related variability time scales. In October 2003, the situation is also rather complex though the prewhitening indicates that frequencies at about 0.35-0.4 d$^{-1}$ dominate the power spectrum. We note also that the same analysis performed on the DIBs did not reveal significant variability due to normalization errors that could disturb our variability analysis. 

\paragraph{H$\beta$.}
This line displays also a variable behaviour, but only at a rather low level. The detection of this variability is most significant for the October 2003 data set, even though data of better quality were obtained during the September 2002 campaign. In October 2003, the variable part of the line could be divided into a red and a blue side. The red part seems to be ruled by a combination of frequencies of about 0.14 and 0.35 d$^{-1}$ (or their $1-\nu$ aliases), or by a frequency of about 0.44 d$^{-1}$. In the blue part, the possible frequencies are close to 0.3 or 0.35 d$^{-1}$ (or their $1-\nu$ aliases). Even though the frequencies of about 0.35 d$^{-1}$ appearing for the \ion{He}{ii} $\lambda$ 4686 and H$\beta$ lines could be related to a common physical process, we did not obtain sufficiently consistent results to give a detailed description of the variable behaviour of these lines.
 
\begin{table}
\caption{Observing runs used for the line profile variability study of HD\,14434. \label{runs3}}
\begin{center}
\begin{tabular}{l c c c c}
\hline
\hline
 Mission & Number of & Line & $\Delta$T & $\Delta\nu_\mathrm{nat}$\\
 & spectra & & (d) & (d$^{-1}$)\\
\hline
Sept. 2002 & 11 & \ion{He}{ii} $\lambda$ 4686 & 14.936 & 0.067\\
	& & H$\beta$ & & \\
Oct. 2003 & 19 & \ion{He}{ii} $\lambda$ 4686 & 17.891 & 0.056\\
	& & H$\beta$ & & \\
\hline
\end{tabular}
\end{center}
\end{table}

\section{Discussion \label{disc}}
At this stage, it is interesting to recall the different time scales which emerged from our line profile analyses for the three stars. Table\,\ref{rec} provides the frequencies found (as well as the corresponding periods) for the different lines, and for each data set. Because of the ambiguity between aliases ($\nu$ and 1$-\nu$), they are both given in most cases, with the one of lowest frequency in brackets.

\begin{table*}
\caption{Summary of the frequencies emerging from our line profile analyses of HD\,192281, HD\,14442 and HD\,14434. The frequencies are given with their 1$-\nu$ aliases in brackets. When two frequencies are quoted the aliases are not specified. When more than one value is provided for a line of an observing run, it indicates the results obtained for different wavelength regions of the same line, from the blue to the red. ND means that `no data' were available for that specific line and NV stands for `no significant variability'. \label{rec}}
\begin{center}
\begin{tabular}{l c c c c c c c}
\hline
\hline
	&	& \multicolumn{2}{c}{HD\,192281} & \multicolumn{2}{c}{HD\,14442} & \multicolumn{2}{c}{HD\,14434} \\
 Observing run & Line &  $\nu$ & P & $\nu$ & P & $\nu$ & P \\
 & & (d$^{-1}$) & (d) & (d$^{-1}$) & (d) & (d$^{-1}$) & (d) \\
\hline
September 1998 	& \ion{He}{ii} $\lambda$ 4686 & 0.63 (0.37) & 1.59 (2.70) & ND &  ND & ND & ND \\
 & & & & & & & \\
Summer 1999 & \ion{He}{ii} $\lambda$ 4686 & 0.70 (0.33) & 1.42 (3.03) & 0.66 (0.33) & 1.52 (3.03) & ND & ND \\
	&	&  &  & 0.59 (0.41) & 1.69 (2.44) & & \\
	& H$\beta$ & 0.67 (0.33) & 1.49 (3.03) & NV & NV & ND & ND\\
 & & & & & & & \\
September 2000 & \ion{He}{ii} $\lambda$ 4686 & ND & ND & 0.62 (0.38) & 1.61 (2.63) & ND & ND\\
	&	&  &  & 0.69 (0.31) & 1.45 (3.23) & & \\
	&	&  &  & 0.76 (0.24) & 1.32 (4.17) & & \\
	& H$\beta$ & ND & ND & 0.65 (0.35) & 1.54 (2.86) & ND & ND \\
 & & & & & & & \\
September 2001 & \ion{He}{ii} $\lambda$ 4686 & 0.66 (0.35) & 1.52 (2.86) & 0.61 (0.40) & 1.64 (2.50) & ND & ND\\
	&	& 0.64 (0.36) & 1.56 (2.78) &  & & & \\
	& H$\beta$ &  0.14 \& 0.46 & 7.14 \& 2.17 & 0.61 (0.39) & 1.64 (2.56) & ND & ND \\
 & & & & & & & \\
September 2002 & \ion{He}{ii} $\lambda$ 4541 & NV & NV & 0.75 (0.26) & 1.33 (3.85) & NV & NV \\
	&	&  &  & 0.63 (0.37) & 1.59 (2.70) & & \\
	& \ion{He}{ii} $\lambda$ 4686 & 0.64 (0.36) & 1.57 (2.78) & 0.63 (0.37) & 1.59 (2.70) & 0.91 & 1.10 \\
	&	& 0.65 (0.36) & 1.54 (2.78) & 0.66 (0.34) & 1.52 (2.94) & 0.92 & 1.09 \\
	&	& 0.67 (0.33) & 1.49 (3.03) & 0.77 (0.23) & 1.30 (4.35) & 0.70 \& 0.91 & 1.43 \& 1.10 \\
	&	&  &  & 0.61 (0.39) & 1.64 (2.56) & & \\
	& H$\beta$ & NV & NV & 0.65 (0.35) & 1.54 (2.86) & NV & NV \\
	&  &  &  & 0.77 (0.23) & 1.30 (4.35) & & \\
	&  &  &  & 0.64 (0.36) & 1.56 (2.78) & & \\
 & & & & & & & \\
October 2003 & \ion{He}{ii} $\lambda$ 4686 & 0.67 (0.33) & 1.49 (3.03) & 0.61 (0.39) & 1.64 (2.56) & 0.15 \& 0.60 & 6.67 \& 1.67 \\
 & & 0.59 (0.41) & 1.69 (2.44) &  &  & 0.65 \& 0.82 & 1.54 \& 1.22\\
 & &  &  &  &  & 0.66 \& 0.83 & 1.52 \& 1.20 \\
 & H$\beta$ & 0.13 \& 0.26 & 7.69 \& 3.85 & 0.57 (0.41) & 1.75 (2.43) & 0.44 & 2.27 \\
 & &  &  &  &  & 0.14 \& 0.74 & 7.14 \& 1.35 \\
 & &  &  &  &  & 0.34 \& 0.74 & 2.94 \& 1.35 \\
\hline
\end{tabular}
\end{center}
\end{table*}

It appears from Table\,\ref{rec} that some variability time scales are found in different data sets of a same star. This is obvious for HD\,192281 and HD\,14442. This recurrence in some frequencies suggests that a rather stable process is at work in the variability we observe. However, the case of HD\,14434 is much less conclusive. Indeed, our data sets did not reveal any obvious time scale which could be responsible for a large part of the variability detected for the \ion{He}{ii} $\lambda$ 4686 line, and to some extent for H$\beta$. This points to a more complex situation than for HD\,192281 and HD\,14442. Nevertheless, probably the most important result arising from our line profile analysis of HD\,14434 is the absence of variations on time scales of several hours. This is especially important in the framework of scenarios where non-radial pulsations may play a role (see Sect.\,\ref{other}). The non-detection of short time scale variations suggests that pulsations do not have a major impact in this Oef star, contrary to what was put forward by the study of BD\,+60$^\circ$\,2522 (Paper I), $\zeta$\,Pup (Reid \& Howarth \cite{RH}) and $\lambda$\,Cep (de Jong et al.\,\cite{deJ}).\\

In addition, there are several points that have to be considered in a model for the variability discussed in this study. First, the stars are rapid rotators and display a double peaked \ion{He}{ii} $\lambda$ 4686 emission line. Second, the `periodic' variability observed during a specific observing run affects only part of the line profile. Third, the variability pattern changes from one epoch to the other and more specifically, the wavelength range displaying the periodic behaviour can shift from one wing of the line to the other. Finally, the variability level changes as well from one oberving run to the next.

\subsection{Rotational modulation}

Since Oef stars are fast rotators, a straightforward question is whether a stable variability time scale as observed in the case of HD\,192281 and HD\,14442 could be linked to stellar rotation. To address this issue, we must first derive some constraints on the rotational period of these stars. An upper limit on the rotational period can be obtained from the observed projected rotational velocity.
\begin{equation}
\label{pmax}
P_\mathrm{max}
=
\frac
{2\,\pi\,R_\mathrm{eq}}
{V_\mathrm{rot}\,\sin{i}}
\end{equation}
where $R_\mathrm{eq}$ is the stellar radius at the equator (see Table\,\ref{gen}).\\
A lower limit can be evaluated as the inverse of the critical rotational frequency, beyond which the centrifugal acceleration overwhelms the effective gravity acceleration.
This condition yields the relation (\ref{pmin})
\begin{equation}
\label{pmin}
P_\mathrm{min}
= 
\frac
{2\,\pi\,R_\ast^{3/2}}
{\left(
G\,M_\ast\,(1-\Gamma)
\right)^{1/2}}
\end{equation}
where $G$ is the gravitational constant, $M_\ast$ the mass of the star, and $\Gamma$ the ratio of the luminosity of the star to the Eddington luminosity. This parameter was evaluated with the following relation given by Lamers \& Leitherer (\cite{LL})
\begin{equation}
\label{gam}
\Gamma
=
7.66\,10^{-5}\,\sigma_\mathrm{e}\,
\left(
\frac
{L_\ast}
{L_\odot}
\right)
\left(
\frac
{M_\odot}
{M_\ast}
\right)
\end{equation}
where all parameters have their usual meaning, with the electron scattering coefficient $\sigma_\mathrm{e}$ taking the value 0.34 cm$^2$\,g$^{-1}$.\\

\subsubsection{HD\,192281 and HD\,14442}

Let us first consider the case of HD\,192281. We adopt $V_\mathrm{rot}\,\sin{i}=$270 km\,s$^{-1}$ as obtained by Conti \& Ebbets (\cite{CE}). With the values taken from Leitherer (\cite{Lei}) for the radius (15 R$_\odot$), the luminosity (10$^{5.94}$ L$_\odot$) and the stellar mass (70 M$_\odot$), we find that the rotation period should be in the range from 0.98 to 2.81 d. Those boundaries correspond respectively to frequencies of 1.02 and 0.36 d$^{-1}$.\\
We see in Table\,\ref{rec} that most of the periods range between 1.40 and 1.59 d (with aliases between 2.70 and 3.44 d). It is noticeable that most of the peaks of our periodograms lie so close to each other, and that the results are similar for several epochs. From the limits discussed hereabove, we find that most probably only the group of periods near 1.5 d could possibly be related to the rotational period.\\

The case of HD\,14442 is even more interesting. By the use of equations (\ref{pmax}) to (\ref{gam}), we calculated the constraints on the rotational period. We adopted a luminosity of $\log$(L/L$_\odot$)\,=\,5.62 intermediate between the two values given by Kendall et al. (\cite{Ken}). For the radius, we chose the value 12.4 R$_\odot$ corresponding to the luminosity given hereabove, and the T$_\mathrm{eff}$ from Table\,\ref{gen}. Kendall et al. (\cite{Ken}) estimated the mass of HD\,14442 by combining the stellar radius with atmospheric parameters, and by plotting its bolometric magnitude versus T$_\mathrm{eff}$ on a HR-diagram and interpolating with theoretical evolutionary tracks. They found two values of respectively about 44 M$_\odot$ and 54 M$_\odot$. We use an intermediate value of 49 M$_\odot$. We find limits on the rotational period of P$_\mathrm{min}$ $=$ 0.82 d and P$_\mathrm{max}$ $=$ 2.29 d. These periods correspond to frequencies of $\nu_\mathrm{max}$ $=$ 1.22 d$^{-1}$ and $\nu_\mathrm{min}$ $=$ 0.44 d$^{-1}$.\\
As shown in Table\,\ref{rec}, the results are quite consistent from one line to the other (\ion{He}{ii} $\lambda$ 4686 and H$\beta$), and from one epoch to the other. The first group of periods lies between 1.52 and 1.69 d, and the second one between 2.44 and 3.03 d. Since the rotational period of HD\,14442 should range between 0.82 and 2.29 d, only the first group of periods could possibly be related to a rotational modulation.\\ 

\begin{figure}
\begin{center}
\resizebox{8.5cm}{4.0cm}{\includegraphics{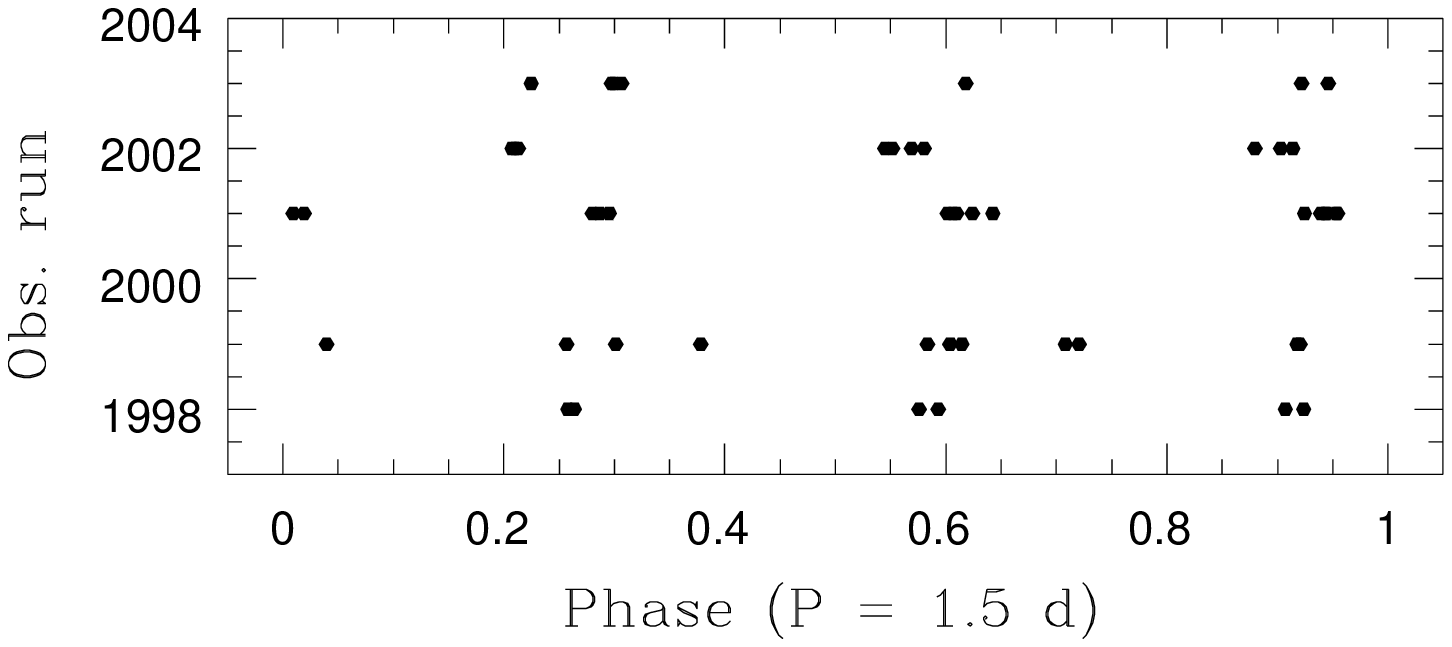}}
\end{center}
\begin{center}
\resizebox{8.5cm}{4.0cm}{\includegraphics{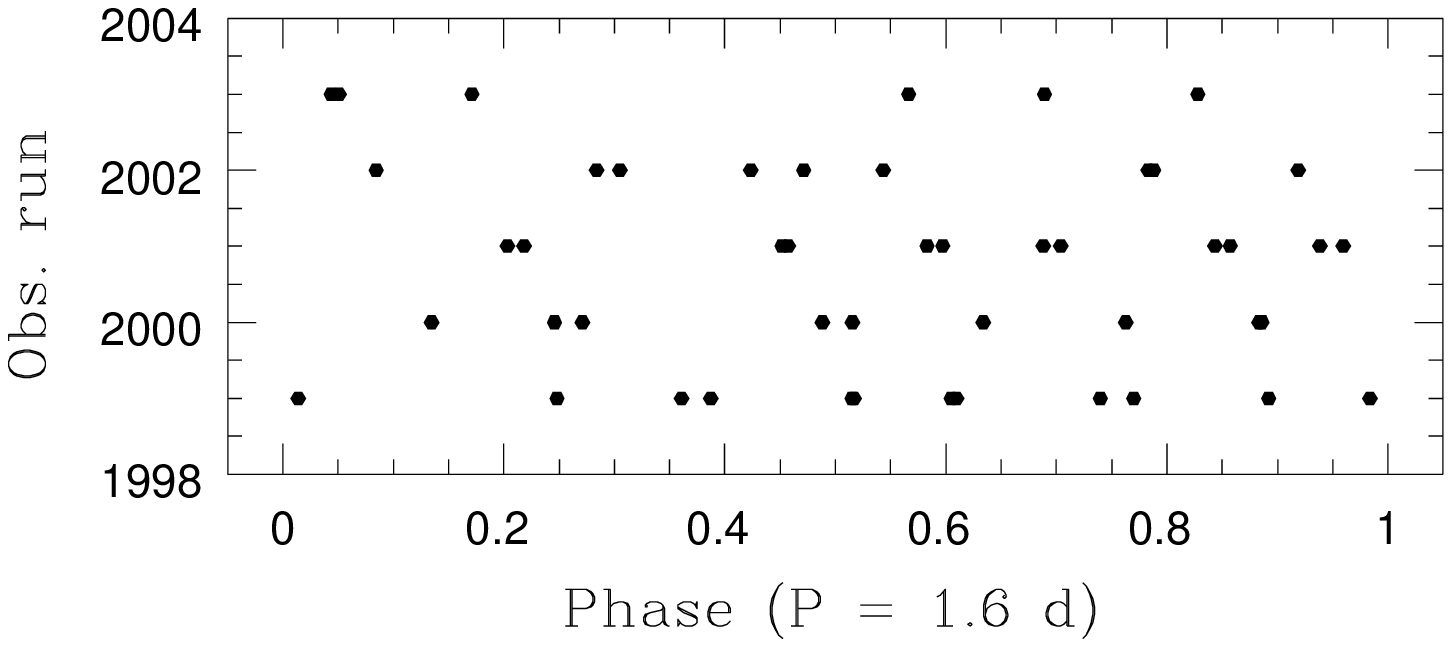}}
\end{center}
\caption{Phase coverage of our data sets considering the proposed periods of 1.5 and 1.6 d respectively for the variability of HD\,192281 (upper panel) and HD\,14442 (lower panel). The zero time for phase calculation is arbitrarily chosen at HJD 2\,450\,000 for both stars. Each horizontal series of points corresponds to an observing run whose date is indicated on the vertical axis labelled in years.\label{phase}}
\end{figure}

Therefore, assuming that the line profile variability discussed here is indeed related to rotation, we tentatively infer rotational periods of about 1.5 and 1.6 d for HD\,192281 and HD\,14442 respectively. The uncertainties on these periods would be of the order of 0.1 d. From the projected rotational velocities and the stellar radii (see Table\,\ref{gen} and Sect.\,3), we can infer inclination angles of 30$-$35$^\circ$ and 41$-$47$^\circ$ for the rotational axis of HD\,192281 and HD\,14442 respectively.\\
An important issue to address is whether our data sets provide a sufficient sampling of the periods that apparently rule the line variability in these two stars. Fig.\ref{phase} shows the phase coverage of all the data sets used for this study. These plots were obtained assuming periods of 1.5 and 1.6 d respectively for HD\,192281 and HD\,14442. In the case of HD\,192281, we see that most points cluster around three phases. This is due to the fact that a period of 1.5 d cannot be fully sampled from a single observatory on the ground, whatever the number of spectra obtained. However, as stated in the previous paragraph, the confidence interval for this period ranges from 1.4 to 1.6 d. Except around the central value of this domain, the phase coverage appears to be satisfactory. The case of HD\,14442 is less problematic. With a period of about 1.6 d, we are not confronted to the same difficulty as with HD\,192281. For each mission, the phase coverage is satisfactory, except for values close to the lower boundary of the confidence interval. For both stars, the phase coverage for the selected periods becomes satisfactory when all our data sets are combined.\\

Fast rotation had been suggested to produce so-called wind compressed zones (Bjorkman \& Cassinelli \cite{BC}) near the stellar equator. However, more recent models suggest that non radial radiative forces prevent the formation of such an equatorial compressed wind (Owocki at al.\,\cite{OCG}) and tend rather to focus the wind towards the poles. The non-spherical shape of a rapidly rotating star will also lead to enhanced mass loss at the pole through gravity darkening (Maeder \cite{Mae}). Still, an equatorially confined wind could be produced by the effect of a stellar magnetic field.\\ 
The rotational broadening of absorption lines in the spectra of O stars renders a direct measurement of their magnetic field extremely difficult. Nevertheless, dipolar magnetic fields of 360 G and 1100 G have been measured for $\beta$\,Cep and $\theta^1$Ori\,C respectively using spectropolarimetric techniques (Donati et al.\, \cite{Don1}, \cite{Don2}). These magnetic field strengths are sufficient to confine the stellar winds into the region of the magnetic equator. Theoretical models for such confined corotating structures have been developed by Babel \& Montmerle (\cite{BM}) and very recently by ud-Doula (\cite{ud}). The latter author considered the case of aligned magnetic and rotational axes. He found that a corotating wind structure forms if the rotational velocity reaches a significant fraction of the critical velocity. According to ud-Doula (\cite{ud}), the corotating structure in the wind consists in a density enhancement of the plasma with a lower outflow velocity. This density enhancement tends to increase as the rotational velocity approaches the critical value, and increases also as the magnetic field strength increases. In our case, the rotational velocities reach several tenths of the critical velocities and a magnetic field of a few hundred Gauss could be sufficient to trigger a corotating wind structure. Such a corotating structure might explain the double-peaked morphology of the \ion{He}{ii} $\lambda$ 4686 emission in Oef stars. Moreover, if the magnetic field is tilted with respect to the rotational axis, we expect the corotating confined wind to produce a rotational modulation of the spectral lines formed in the wind. Such an `oblique magnetic rotator' model was put forward e.g. by Moffat \& Michaud (\cite{MM}), Stahl et al. (\cite{Sta}) and Rauw et al. (\cite{Rauw}) to explain the line profile variations of $\zeta$\,Pup, $\theta^1$Ori\,C and HD\,192639 respectively.\\
The inclination of an oblique corotating wind structure with respect to our line of sight is expected to vary with a period equal to the rotational period. This general scenario could lead to a large number of variability patterns depending on the inclination angle of the rotation axis on the one hand, and the tilt angle between the magnetic and the rotational axes on the other hand. The larger the tilt angle, the larger the contrast between the observed configuration of the confined corotating wind structure at different phases. The radial extent of the corotating wind would depend on the wind velocity and the magnetic field strength. Therefore, lines formed over a large radius would be less affected by the rotational modulation than those formed in the innermost part of the wind where the effect would be most pronounced.\\
The oblique magnetic rotator model might account for the recurrence of a stable clock in the line profile variability. However, as pointed out above, the variability with this time scale only occurs over a limited range of wavelength and shifts from the red wing to the blue wing (and vice versa) from one year to the next. If our data provide a good sampling of the phenomenon, this fact cannot easily be explained by the simple tilted magnetic rotator model described hereabove.

\subsubsection{HD\,14434}
In the case of HD\,14434, stellar rotation does not appear a priori as a good candidate to interpret the line profile variations that we observed. Our data did not allow us to clearly identify frequencies responsible for the variable behaviour investigated in Sect.\,\ref{var3}. However, we should notice that the search for a reccurent variability time scale requires more observations of this star.
 
\subsection{Other scenarios \label{other}}
Beside the rotational modulation discussed above, other factors are likely to influence the variability of the objects studied in this paper. Indeed, although the main time scales of the variability of HD\,192281 and HD\,14442 are broadly compatible with the rotation period, some aspects do not fit this interpretation. This is even more true for the cases of HD\,14434 and BD\,+60$^\circ$\,2522 (Paper I).\\

First, in the framework of the simulations of ud-Doula (\cite{ud}), it is shown that for large wind confinement parameters, i.e. large magnetic field strengths, the density structure near the surface becomes quite complex and undergoes a complicated variability behaviour. The occurence of this phenomenon at distances from the star corresponding to the formation region of the \ion{He}{ii} line could have a strong effect on its shape.\\  
Second, we know that the best studied Oef stars ($\zeta$\,Pup and $\lambda$\,Cep) undergo non-radial pulsations (see Reid \& Howarth \cite{RH}; de Jong et al.\,\cite{deJ}). Cranmer (\cite{Cran}) considered the propagation of non-radial pulsations (NRPs) into accelerated stellar winds, and showed how this could lead to large scale structures likely able to generate significant line profile modulations. The study of NRPs in rotating early-type stars is also considered by Townsend (\cite{Tow}). According to the latter author, rotation compresses pulsational activity towards the equator. NRPs and probably multimodal pulsations are believed to trigger mass loss enhancements (or depletions) possibly responsible of the apparition of structures in the stellar wind. Moreover, the combined effects of rotation and strong magnetic fields on NRPs is not well understood and can lead to complex cases like the so called `off-axis modes' encountered for Ap stars for which the axis of pulsation does not coincide with the rotation axis (Townsend \cite{Tow}). These considerations could provide the ingredients for an alternative scenario. The interplay between NRPs, rotation and the stellar wind were proposed to influence the variability behaviour of BD\,+60$^\circ$\,2522 (Paper I), because of the lack of stability in the variability pattern in that case. However, at least in the case of HD\,14434, a series of spectra obtained with about 30 minutes separation over more than 6 hours in October 2003 did not reveal significant short-term variability for any of the spectral lines in our wavelength domain. This result argues against a major role of pulsations at least for HD\,14434, but high signal-to-noise ratio and high spectral resolution data, as well as a better sampling, are required to confirm this result.\\
Third, some variability should also be expected due to outmoving clumps in the wind of O stars. In the case of $\zeta$\,Pup, spectral substructures similar to those observed in WR spectra are probably the consequence of excess emission from clumps (Eversberg et al.\,\cite{ELM}). The variation across the line can reach an amplitude of about 5\,\% of the total line intensity. However, the substructures in $\zeta$\,Pup vary on time scales of a few hours only, which is too short to suit the variability time scales revealed by the study of our sample of Oef stars and is again at odds with the results of our intense monitoring of HD\,14434.\\
Finally, we should also consider the possibility of the propagation of a clump (as discussed in the previous paragraph) across a putative confined structure within the wind (as discussed in the previous section) or a density wave. Such a scenario could lead to modulations in the profiles of the lines formed within the confined wind. The interaction of such perturbations with a corotating structure should be considered in future theoretical developments, in order to establish its impact on line profiles. 

\section{Conclusions and future work \label{conc}}
We have performed a spectral analysis of the Oef stars HD\,192281, HD\,14442 and HD\,14434. The double peaked structure of the emission component of the \ion{He}{ii} $\lambda$ 4686 line of these three objects is obvious, and points to a possible corotating structure covering a significant fraction of the emitting region of this line in the stellar wind (Conti \& Leep \cite{CL}). We determined spectral types (respectively O5(ef), O5.5ef and O6(ef) for the three stars) in agreement with former classifications. Our radial velocity measurements did not reveal any evidence of binarity on time scales of several days, or from one year to the next. The peculiar velocity of HD\,192281 does not support a putative runaway status for this star.\\

A detailed line profile variability analysis was performed on the strongest lines in the blue spectra of these stars. We have shown that the \ion{He}{ii} $\lambda$ 4686 line of HD\,192281 and HD\,14442 displays the most significant variability, with time scales which are consistently found from one epoch to the other for both stars. We find however that the structure of the profile variability changes with time. The `periodic' modulations of the profile, which are confined in a narrow wavelength domain, switch forth and back between the blue and the red wings, but never leave the central absorption component of the line\footnote{Spectral modulations on a time scale compatible with the rotation period, and located in the central region of the \ion{He}{ii} $\lambda$ 4686 line, were also observed in the case of $\zeta$\,Pup (Eversberg et al.\,\cite{ELM}).}. The variability level itself changes also depending on the data set. Assuming that the periodicity of this variability is ruled by the rotation of the star, we propose rotational periods of 1.5 $\pm$ 0.1 and 1.6 $\pm$ 0.1 d for HD\,192281 and HD\,14442 respectively. HD\,14434 displays the most complex behaviour on a yet undefined time scale. However, this star does not display any variations on time scales of a few hours, arguing against a pulsational origin for the observed line profile variations. The H$\beta$ line displays lower level variations. In the case of HD\,14442, the variations of H$\beta$ are correlated with those observed in \ion{He}{ii} $\lambda$ 4686.\\

In the framework of the most recent theoretical results concerning the magnetohydrodynamics of magnetic field-aligned rotating winds (ud-Doula\,\cite{ud}), we tentatively propose a scenario to explain some aspects of the observed line profile variability of HD\,192281 and HD\,14442. In this scenario, the interaction of the magnetic field with the wind plasma leads to the confinement of a significant part of the stellar wind in the vicinity of the magnetic equator plane. While several aspects of the observed variability behaviour could be explained by a corotating structure resulting from this interaction, we note that this scenario cannot explain the wavelength confinement of the variability pattern nor the shift of this pattern from the blue to the red from one year to the next, if our sampling is adequate. The wavelength confinement of the variability around zero velocity we describe in this study suggests that the structures embedded in the stellar wind have a limited spatial extent, and the change in the part of the line which suffers this variability points to the fact that these structures are not static. Moreover, like in the case of BD\,+60$^\circ$\,2522 (Paper I), the variations suffered by the line profiles of HD\,14434 can not be explained by this scenario.\\
The description of the variability of Oef stars constitutes a challenge for theoretical modelizations. Further magnetohydrodynamical models including the interplay between NRPs, magnetic fields and rotation need to be developed in order to clarify the complex behaviour of the lines in the spectra of Oef stars. On the other hand, from the observational point of view, a similar study of $\zeta$\,Pup and $\lambda$\,Cep should be envisaged to check whether the variability time scales and the variation patterns reported in the literature are stable over longer time scales. We note also that the study of these stars should benefit strongly from the use of a wide-band spectrograph, as well as a denser time sampling.

\acknowledgement{We wish to express our thanks to Dr. Jean Manfroid for taking some of the spectra of the September 2000 campaign. We are greatly indebted to the Fonds National de la Recherche Scientifique (Belgium) for multiple assistance including the financial support for the rent of the OHP telescope in 1999, 2000 and 2002 through contract 1.5.051.00 "Cr\'edit aux chercheurs" FNRS. The travels to OHP for the observing runs were supported by the Minist\`ere de l'Enseignement Sup\'erieur et de la Recherche de la Communaut\'e Fran\c{c}aise. This research is also supported in part by contracts P4/05 and P5/36 "P\^ole d'Attraction Interuniversitaire" (Belgian Federal Science Policy Office) and through the PRODEX XMM-OM Project. We would like to thank the staff of the Observatoire de Haute Provence for their technical support during the various observing runs. The SIMBAD database has been consulted for the bibliography.}

%\listofobjects
\end {document}